\documentclass[aps,prl,superscriptaddress,amsmath,amssymb,floatfix,twocolumn]{revtex4-1}
\usepackage{times}
\usepackage{graphicx}
\usepackage{subfigure}
\usepackage{color}
\usepackage{epstopdf}

\begin{document}

\title{Mott transition in Modulated Lattices and Parent Insulator of (K,Tl)$_{\rm y}$Fe$_{\rm x}$Se$_{\rm 2}$ Superconductors}

\author{Rong Yu}
\affiliation{Department of Physics \& Astronomy, Rice University, Houston, Texas 77005}
\author{Jian-Xin Zhu}
\affiliation{Theoretical Division, Los Alamos National Laboratory,
Los Alamos, New Mexico 87545}
\author{Qimiao Si}
\affiliation{Department of Physics \& Astronomy, Rice University, Houston, Texas 77005}

\begin{abstract}
The degree of electron correlations remains a central issue in the iron-based
superconductors. The parent iron pnictides  are antiferromagnetic,
and their bad-metal behavior has been interpreted in terms of proximity to a Mott transition.
We study such a transition in multi-orbital models on modulated lattices containing
an ordered
pattern of iron vacancies,
using a slave-rotor method. We show
that the ordered vacancies lead to a band-narrowing,
which pushes
the system to the Mott insulator side.
This effect is proposed to
underlie the insulating behavior observed in the parent compounds of the newly discovered
(K,Tl)$_{\rm y}$Fe$_{\rm x}$Se$_{\rm 2}$ superconductors.
\end{abstract}

\maketitle

{\it Introduction.~} Superconductivity in the layered iron pnictides and chalcogenides
 occurs
 near antiferromagnetically-ordered parent compounds
 ~\cite{Kamihara_FeAs,Zhao_Sm1111_CPL08,Cruz}.
In their paramagnetic phase, these parent materials have a large
 electrical resistivity corresponding to an in-plane carrier mean-free-path
 on the order of the inverse Fermi wavevector.
 They also show a strong reduction of the Drude weight~\cite{Qazilbash},
 and temperature-induced spectral weight transfer that extends to high energies
(on the eV order)~\cite{hu,boris,yang}.
Such bad-metal behavior is characteristic of metallic systems in proximity
 to a Mott transition~\cite{Si, Si_NJP,Kupetov}

\begin{figure}[b!]
\centering\includegraphics[scale=0.28
,bbllx=80pt,bblly=0pt,bburx=760pt,bbury=980pt
]{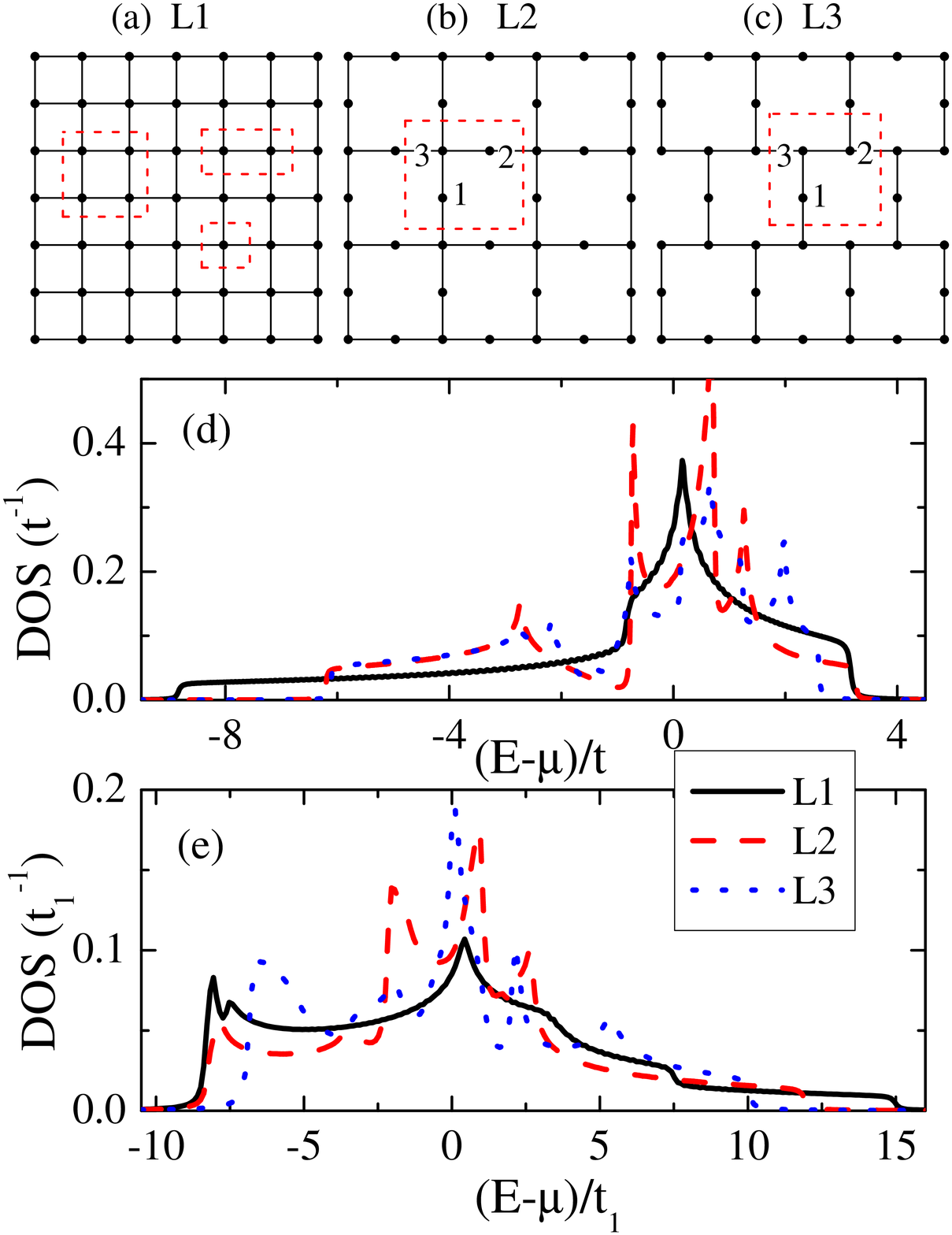}
\caption{(Color online)
a) Square lattice, L1. Different clusters are used and will be labeled as
$N_c=1$,$2$, and $4$, respectively; b) Modulated $2\times2$ square lattice, L2. The enclosed sites $1$,$2$,$3$ form the basis of the unit cell;
c) Another modulated $2\times2$ lattice, L3 (corresponding to a $4\times2$ superstructure in the FeSe plane). Also shown are the bare density-of-states (DOS) for the one-orbital model ($t=t'=1$) (d) and the two-orbital model (e).
}
\label{fig:1}
\end{figure}

Recently, superconductivity has been discovered in a new family of iron-based compounds
K$_{\rm y}$Fe$_{\rm 2}$Se$_{\rm 2}$ \cite{Guo}
and related (K,Tl)Fe$_{\rm x}$Se$_{\rm 2}$ \cite{Fang}.
In these compounds the maximal superconducting transition temperature is comparable to that of the 122 iron pnictides. Similarly to the pnictides, the superconductivity occurs
close to an antiferromagnetically ordered state~\cite{Fang}.
At the same time, these materials are unique in several aspects.
Both the angle-resolved photoemission (ARPES) experiments \cite{Zhang_Feng,Qian_Ding,Mou} and LDA calculations~\cite{Shein} show that the Fermi surface has only electron pockets.
The absence of hole Fermi pockets is unique among the iron based superconductors,
 raising hope for major new and general insights to be gained from studying these materials.
Equally important,
the Fe vacancies may form ordered patterns when the Fe content $x\lesssim 1.6$
as suggested by various experiments~\cite{Fang,Wang,Bao1102_0830}.
Furthermore, there are parent compounds which are
insulating~\cite{Fang,Wang1101_0789}.
The control parameter that tunes the (K,Tl)Fe$_{\rm x}$Se$_{\rm 2}$
 system from superconducting to
insulating is the Fe composition $x$, and $x=1.5$ is the primary candidate composition
for a parent compound. There is evidence \cite{Fang,Wang} that in
(K,Tl)Fe$_{\rm 1.5}$Se$_{\rm 2}$ the Fe vacancies form regular patterns possibly
as illustrated in either Fig.~\ref{fig:1}(b) or Fig.~\ref{fig:1}(c).
The in-plane electrical resistivity  is about two orders of magnitude
larger than that of the parent iron pnictides at room temperature,
and it further increases exponentially as temperature is lowered.
The insulating behavior is also manifested in the optical conductivity \cite{ZGChen},
 which is strongly suppressed below about 0.7 eV.
Because of experimental indications that the (K,Tl) content is also variable,
 we will in
 the following refer to these systems
 as (K,Tl)$_{\rm y}$Fe$_{\rm x}$Se$_{\rm 2}$.

In this Letter, we propose that the parent (K,Tl)$_{\rm y}$Fe$_{\rm 1.5}$Se$_{\rm 2}$
 is a Mott insulator arising from a correlation effect that is enhanced by
  the Fe vacancies. We  describe the ordered Fe vacancies in terms
  of a modulated lattice, and introduce a two-orbital model with two electrons
  per Fe site to capture their electronic structure.
We use a slave-rotor method to show that a Mott transition
exists in this model even though there are an even number of electrons
per site (and per unit cell).
We find that the interaction strength for the Mott transition largely tracks the electronic bandwidth.
In other words, ordered Fe vacancies enhance the tendency towards Mott localization as a result of a kinetic-energy reduction. Such Fe vacancies, therefore, have a similar effect as a lattice
expansion, which we have previously discussed as responsible for the
Mott insulating behavior in La$_2$O$_3$Fe$_2$Se$_2$~\cite{Zhu_prl10}.
Our considerations of the interaction effects are realistic,
given that the {\it ab initio} calculations using
density-functional theory \cite{Cao} show that the
3d bands
of TlFe$_{\rm 1.5}$Se$_{\rm 2}$
are narrower than those of
TlFe$_{\rm 2}$Se$_{\rm 2}$.

{\it Modulated lattices and kinetic-energy reduction.~}
We will consider a square lattice (L1, Fig.~\ref{fig:1}(a)),
a modulated square lattice consisting of $2\times2$ plaquettes each having its center removed (L2, Fig.~\ref{fig:1}(b)), and another
one corresponding to a triangular lattice of such $2\times2$
plaquettes (L3, Fig~\ref{fig:1}(c)).

The (K,Tl)$_{\rm y}$Fe$_{\rm x}$Se$_{\rm 2}$ system involves all five 3d orbitals.
The ARPES experiments \cite{Zhang_Feng,Qian_Ding,Mou}
show electron pockets near $M$ point and suggest very weak electron-like pockets near $\Gamma$ point. The absence of hole pocket near $\Gamma$ point is largely consistent with
 the {\it ab initio} electronic
 bandstructure calculations using local-density-approximation (LDA)
 for (K,Tl)Fe$_{\rm 2}$Se$_{\rm 2}$
 \cite{Shein,Yan_Xiang,Cao_Dai,Nekrasov,Singh}.
This is
in contrast to the case of iron pnictides, and
is easier to model using a two-orbital tight-binding parametrization.
Correspondingly, we consider a two-orbital model with the degenerate $xz$ and $yz$ orbitals (labeled as orbitals 1 and 2) and $n=2$. Inspired by the considerations in the pnictides case \cite{Raghu,Graser}, we introduce a set of
tight-binding parameters. The parameters are listed in Table~\ref{Table:hopping}, and their meanings can be
 inferred from the dispersion functions specified in Eq.~(\ref{eqn2}). We first fit the LDA bandstructure
 obtained on TlFe$_2$Se$_2$ to this two-orbital
 model and then adjust the tight-binding model parameters so that the Fermi surface still
 has
 only
 electron pockets at
 $n=2$. We notice that
 the Fermi surface size is larger than in the bandstructure calculations, but this suffices for our qualitative considerations of the effect of lattice depletion on the Mott transition.
What is important is that, for our parameters, the Fermi surface comprises only electron pockets near the X points
of the 1-Fe per cell Brillouin Zone.
The bandwidth narrowing for L2 and L3 lattices compared to the L1 lattice is shown in
Fig.~\ref{fig:1}(e).

\begin{table}[h]
\begin{ruledtabular}
\begin{tabular}{cccc|cc}
& intra-orbital & (eV) & & inter-orbital & (eV)\\
\hline
$t_1$ & $t_2$ & $t_3$ & $t_9$ & $t_4$ & $t_{12}$ \\
0.093 & 0.081 & -0.222 & -0.038 & 0.023 & -0.038 \\
\end{tabular}
\end{ruledtabular}
\caption{Hopping parameters of the two-orbital
model.}
\label{Table:hopping}
\end{table}

{\it Two-orbital model and the slave-rotor method.~}
We are now in position to specify our model,
\begin{eqnarray}\label{eqn1}
\mathcal{H}&=& -\sum_{ij,\alpha\beta,\sigma}
t_{ij}^{\alpha \beta} c_{i\alpha\sigma}^{\dagger} c_{j\beta\sigma}
+ \frac{U}{2}\sum_i \left(\sum_{\alpha\sigma} n_{\alpha\sigma}\right)^2
\label{hamiltonian}
\end{eqnarray}
where $c_{i\alpha\sigma}$ annihilates an electron in orbital $\alpha$ and spin $\sigma$ on site $i$ of the Fe lattice.
The first term in Eq.~(\ref{eqn1}) describes the electron hopping,
with orbital dependent hopping amplitudes $t_{ij}^{\alpha \beta}=t_1,
t_2, ...,t_{12}$ yielding $\sum_{k\alpha\beta\sigma}\epsilon^{\alpha\beta}_{k\sigma} c^\dagger_{k\alpha\sigma} c_{k\alpha\sigma}$, where
\begin{eqnarray}\label{eqn2}
\epsilon^{11}_{k} &=& -2t_1\cos k_x -2t_2\cos k_y -4t_3\cos k_x\cos k_y \nonumber\\
& & -4t_9\cos2k_x\cos2k_y, \nonumber\\
\epsilon^{22}_{k} &=& -2t_2\cos k_x -2t_1\cos k_y -4t_3\cos k_x\cos k_y \\
& &-4t_9\cos2k_x\cos2k_y, \nonumber\\
\epsilon^{12}_k &=& \epsilon^{21}_k = -4t_4\sin k_x\sin k_y -4t_{12}\sin 2k_x\sin 2k_y.\nonumber
\end{eqnarray}
Small hoppings $t_9$ and $t_{12}$ between the 5th-nearest neighbors are included to reproduce the two electron pockets at $n=2$.
The second term in Eq.~(\ref{eqn1}) is an on-site Coulomb repulsion.
We focus on the effect of lattice depletion on the Mott transition and will not consider
Hund's coupling and pair-hopping terms
for simplicity. All local interactions are expected to have effects similar to $U$.
In particular, the Hund's coupling will reduce the critical $U$  of the Mott transition;
its effects are readily studied within a slave-spin method~\cite{deMedici05,YuSi10}, and
the results will be reported elsewhere.

\begin{figure}[t!]
\centering\includegraphics[
width=1.0\linewidth
,bbllx=40pt,bblly=240pt,bburx=751pt,bbury=580pt
]{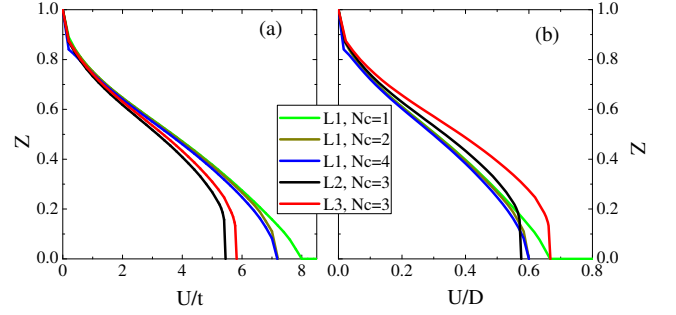}
\caption{(Color online)
  Quasiparticle weight for the one-orbital model in the unmodulated and modulated lattices
  plotted
  as a function of $U/t$ (a) and $U/D$ (b).
}
\label{fig:2}
\end{figure}

We study the model using the cluster slave-rotor mean-field (CSRMF) method~\cite{FlorensGeorges04,ZhaoParamekanti07}.
We introduce an $O(2)$ rotor variable $\theta_i$ and a spinon $f_{i\alpha\sigma}$ on each site,
and write $c_{i\alpha\sigma} = f_{i\alpha\sigma}e^{-i\theta_i}$. Here, $e^{-i\theta_i}$ lowers the rotor angular momentum $L_i$, which corresponds to the charge quantum number. The unphysical states are eliminated by enforcing the constraint $L_i = \sum_{\alpha\sigma} (f^\dagger_{i\alpha\sigma} f_{i\alpha\sigma}-1/2)$ in the enlarged rotor and spinon Hilbert space. By rewriting Eq.~(\ref{eqn1}) using rotor and spinon operators and decoupling the rotor and spinon operators at the mean-field level, we obtain the following two effective Hamiltonians:
\begin{eqnarray}
H_f &=& -\sum_{ij\alpha\beta\sigma} t^{\alpha\beta}_{ij} C_{ij} f^\dagger_{i\alpha\sigma}f_{j\beta\sigma}-(\mu+\lambda) \sum_{i\alpha\sigma} n^f_{i\alpha\sigma},\label{eqn3}\\
H_\theta &=& -2\sum_{ij} t^{\alpha\beta}_{ij} \chi^{\alpha\beta}_{ij} e^{i(\theta_i-\theta_j)} +\frac{U}{2} \sum_{i} L^2_i + \lambda\sum_{i} L_i,\label{eqn4}
\end{eqnarray}
where $C_{ij} \equiv \langle e^{i(\theta_i-\theta_j)} \rangle_\theta$ is the rotor
correlation function that renormalizes the quasiparticle hopping parameters in the presence of interaction,
$\chi^{\alpha\beta}_{ij} \equiv \langle f^\dagger_{i\alpha\sigma}f_{j\beta\sigma}\rangle_f$,
$\mu$ is the chemical potential, and $\lambda$
is a Langrange multiplier to impose the constraint.
To solve these two  Hamiltonians, which still contain interactions among rotors,
we further apply a cluster mean-field approximation.
We exactly diagonalize the rotor Hamiltonian on a finite cluster, and treat the
 influence of the sites outside the cluster as a mean field.
 We decouple $e^{i(\theta_i-\theta_j)}$ into $e^{i\theta_i}\phi_j$ if $i$ belongs
 to the cluster but $j$ is outside. Here $\phi_i\equiv \langle e^{-i\theta_i}\rangle$
 is the local mean-field parameter. Consequently, $C_{ij}$ is factorized as $C_{ij}\approx\phi^*_i \phi_j$ if either $i$ or $j$ is outside the cluster. In practice, Eqs.~(\ref{eqn3}) and (\ref{eqn4})
 are self-consistently solved by iteratively determining
  the mean-field parameters $\phi_i$, $C_{ij}$, and $\chi^{\alpha\beta}_{ij}$.
  The Mott transition is signaled by a vanishing quasiparticle spectral weight $Z = |\phi_i|^2$.

To gain intuition on the role of the lattice modulation,
we will also study  a one-orbital model with nearest-neighbor hopping, $t$, and
next-nearest-neighbor hopping, $t'$. Fig.~\ref{fig:1}(d) illustrates the reduction
of the bandwidth for L2 and L3 lattices  from
that of the L1 lattice, for the case of $t'=t$. A non-zero $t'$ is chosen for two reasons. It avoids
a perfect nesting in the case of $n=1$, which we study below. It also avoids a flat band in the
case of the L2 lattice: when $t'=0$, for the L2 lattice,
there are two dispersive bands with a combined bandwidth of
$4\sqrt{2}t$, and a flat band in the middle.

\begin{figure}[t!]
\centering\includegraphics[
width=1.0\linewidth
,bbllx=30pt,bblly=235pt,bburx=761pt,bbury=585pt
]{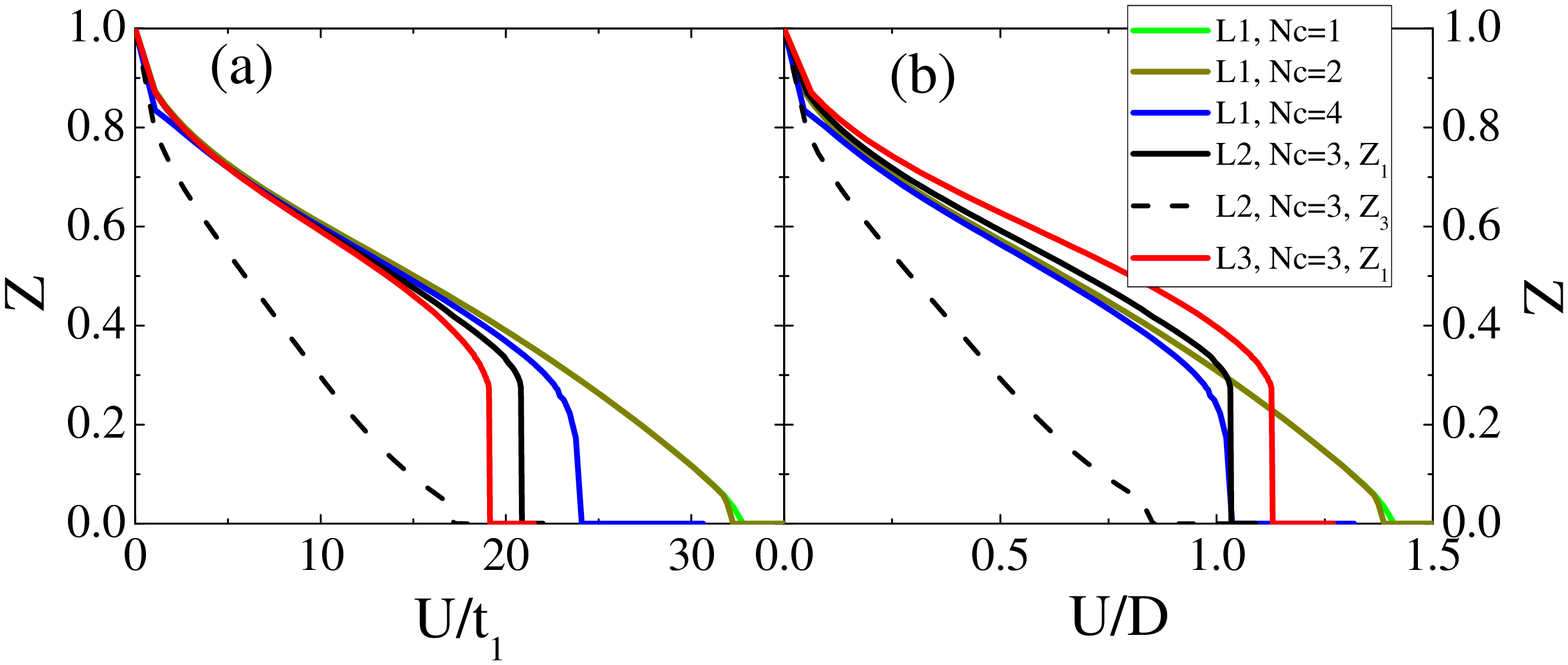}
\centering\includegraphics[
width=1.0\linewidth
,bbllx=30pt,bblly=235pt,bburx=761pt,bbury=585pt
]{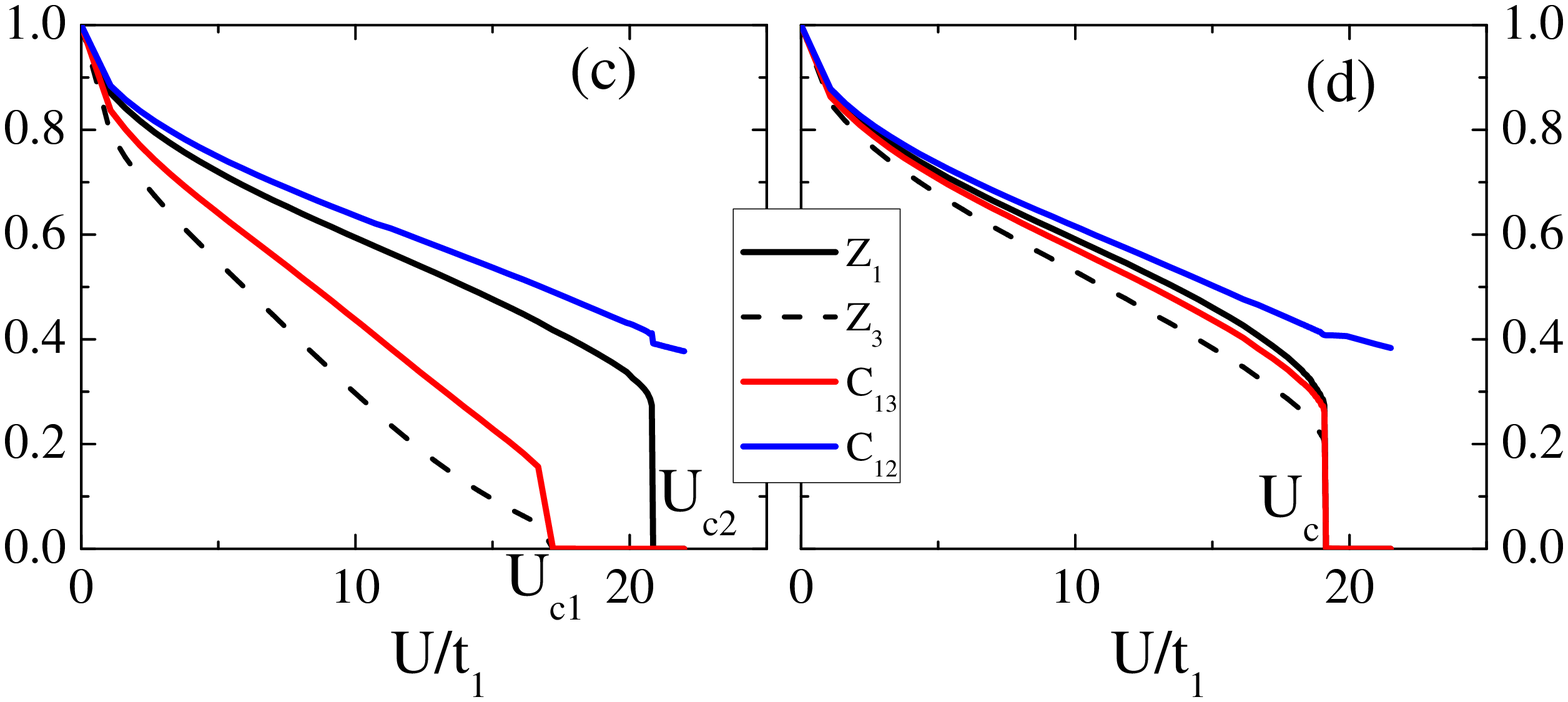}
\caption{(Color online)
Quasiparticle weight $Z$ for the two-orbital model vs. $U/t$ (a) and $U/D$ (b). Also shown are the site-selective $Z_1$ and $Z_3$ and the bond correlations for lattices L2 (c) and L3 (d).
}
\label{fig:3}
\end{figure}

{\it Results for the one-orbital model.~}
We start from the one-orbital case. Because the L2 and L3 lattices involve a $2\times2$ square plaquette as the unit cell, we will carry out our calculations for the lattice L1 with $N_c=4$. The slave-rotor mean-field theory treats the rotor kinetic energy for intra-cluster bonds exactly by diagonalizing the rotor Hamiltonian on the cluster. Hence working with $N_c=4$ gives a better description of the Mott transition than using the single site approximation.
Fig.~\ref{fig:2}(a) shows the renormalized quasiparticle weight, $Z$, as a function of $U/t$. The Mott transition occurs at $U_c$, where $Z$ first goes to zero as $U$ is increased. For the L1 lattice, increasing the cluster size from $N_c=1$ to $N_c=4$ leads to a successive reduction of $U_c$: $U_c\approx 8t$ for $N_c=1$, and $U_c\approx 7.2t$ for $N_c=4$.

On the modulated lattices, we also find a Mott transition. It is seen that 
$U_c\approx5.4t$ for the L2 lattice and $U_c\approx5.8t$ for L3 lattice; 
both are smaller than that of the L1 lattice. 

Fig.~\ref{fig:2}(b) plots the same result, but with $U$ now normalized against the full bandwidth $D$. It is
seen that $U_c/D$ is comparable for all three cases. This clearly illustrates that the reduction of $U_c$ for the modulated L2 and L3 lattices arises from the band-narrowing effect.

\begin{figure}[t!]
\centering\includegraphics[
width=1.0\linewidth,
bbllx=0pt,bblly=200pt,bburx=642pt,bbury=560pt
]{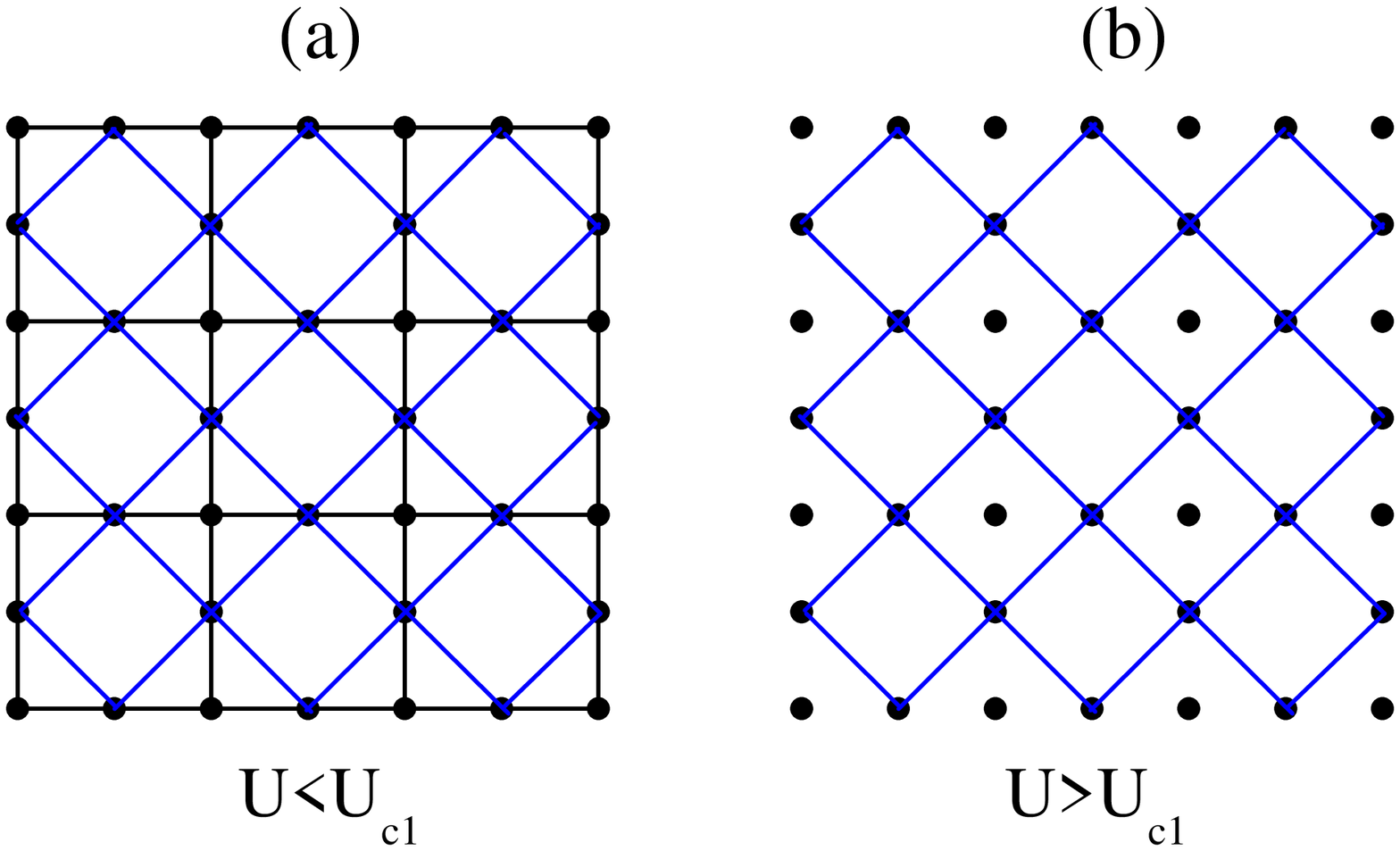}
\centering\includegraphics[
width=1.0\linewidth
,bbllx=30pt,bblly=200pt,bburx=761pt,bbury=585pt
]{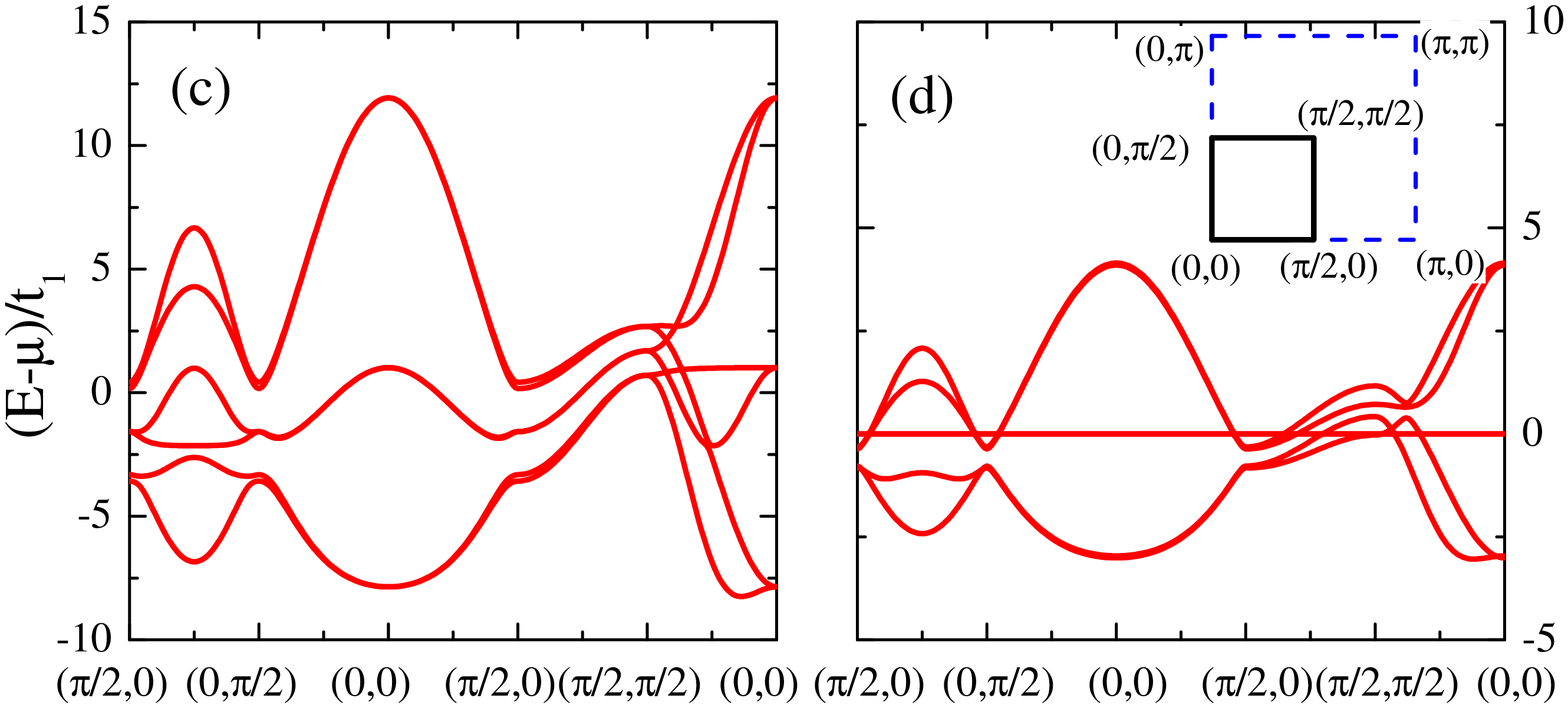}
\caption{(Color online)
Sketch of the renormalized hoppings for the L2 lattice at $U<U_{c1}$ (a) and at $U_{c1}<U<U_{c2}$ (b).
Also shown are the bandstructure of L2 for $U=0<U_{c1}$ (c) and
for $U_{c1}<U=19t_1<U_{c2}$ (d). The bands are shown along the
high-symmetry directions in the Brillouin Zone associated with the
$2\times2$-Fe unit cell, illustrated in the inset in (d).
}
\label{fig:4}
\end{figure}

{\it Results for the two-orbital model.~}
We now turn to the more realistic two-orbital case.
The renormalized quasiparticle weight as a function of $U/t_1$ is shown in Fig.~\ref{fig:3}(a).
It is again seen that the values of $U_c$ for both the L2 and L3 lattices are smaller than that of the L1 lattice with $N_c=4$.

One difference from the toy 1-orbital model is that the hopping parameters in the two-orbital model are highly anisotropic (e.g., $ | t_3/t_2 | \approx 3$), which makes the local environment possibly different from site to site on the modulated lattices. To fully address the influence of the inhomogeneity, we study the quasiparticle weight associated with each site in the cluster
$Z_i=|\langle e^{i\theta_i}\rangle|^2$. For either L1 or L3, we obtain a single Mott transition as in the one-orbital case.
For the L2 lattice, we find two transitions. They are identified by the vanishing of $Z_3$ first at $U_{c1}/t_1\approx17$, and subsequently the vanishing of $Z_1$ (and, equivalently, $Z_2$) at a higher value $U_{c2}/t_1\approx20$.

To understand this, we have in the same figure plotted the bond correlators $C_{12}$ and $C_{13}$.
Between $U_{c1}$ and $U_{c2}$, $C_{13}$ vanishes but $C_{12}$ remains finite. This
makes site $3$ to be unconnected to the rest of the lattice ({\it cf.} Fig.~\ref{fig:4}(b)). We see these explicitly in a plot of the renormalized bandstructure in Fig.~\ref{fig:4}(d): associated with the isolated $3$ sites is a flat band lying exactly on the Fermi level for $U_{c1}<U<U_{c2}$. By contrast, for $U<U_{c1}$,
all sites are connected by hopping terms (Fig.~\ref{fig:4}(a)) and there exists no flat
band (Fig.~\ref{fig:4}(c)).

For the L3 lattice, the geometry prevents the separation of any site from the rest bulk unless all the effective hopping parameters are zero. As a result, there will be only one transition. This is clearly seen in comparing $Z_1$ and $Z_3$ in Fig.~\ref{fig:3}(d).

The quasiparticle weight as a function of $U/D$ is shown in
Fig. \ref{fig:3}(b). It is again seen that $U_c/D$ is comparable for all three cases.
As in the one-orbital case, this illustrates that the reduction of $U_c$ for the modulated lattices
originates from the band-narrowing effect.

{\it Implications for (K,Tl)$_{\rm y}$Fe$_{\rm x}$Se$_{\rm 2}$.~}
Our results imply that the critical interaction strength for the Mott transition will be smaller in
(K,Tl)$_{\rm y}$Fe$_{\rm x}$Se$_{\rm 2}$  than in iron arsenides and 11 iron chalcogenides.
This provides
 the basis for a Mott-insulating state in the parent (K,Tl)$_{\rm y}$Fe$_{\rm x}$Se$_{\rm 2}$, even when
 one assumes the same strength of
Coulomb interaction across all families of iron-based superconductors..

The Mott-insulating nature of the parent (K,Tl)$_{\rm y}$Fe$_{\rm x}$Se$_{\rm 2}$
 is supported by experiments.
As already mentioned,
 the materials for both $x=1.5$ and $x=1.64$ have a large electrical resistivity with an insulating temperature dependence~\cite{Fang,Wang1101_0789}..
 Furthermore, the insulating behavior in the electrical resistivity is already observed in the paramagnetic phase.
Relatedly, the optical conductivity is not only strongly suppressed below about 0.7 eV,
but also small in magnitude. For reference,
the value of the optical conductivity
is comparable to that of the insulating YBa$_{\rm 2}$Cu$_{\rm 3}$O$_{6+x}$
with a slight off-stoichiometry $x=0.2$ ~\cite{Orenstein}.
Finally, magnetic order is known to exist in TlFe$_{\rm x}$Se$_{\rm 2}$
at $x$ close to $1.5$ ~\cite{Haggstrom}. Taken together, these experiments suggest
that the insulating state is of the Mott type.

We note that in compounds with Fe content close to $x=1.6$,
 the ordered vacancies have a different pattern~\cite{Fang,Bao1102_0830}.
However, because we have shown that the Mott localization is a result of
 vacancy-ordering induced band narrowing, our argument will also apply to these systems.
Band narrowing is expected on the ground of general considerations given here,
and can also be seen in
the LDA results ~\cite{Cao}.

To summarize, we have
used a two-orbital model in 1/4-depleted lattices to
demonstrate that
ordered vacancies enhance the tendency for Mott transition, and that this enhancement
originates from a vacancy-induced kinetic-energy reduction. Our qualitative conclusion is
expected to apply to the more realistic five-orbital model. Based on our calculations,
we propose that the insulating parent of the
(K,Tl)$_{\rm y}$Fe$_{\rm x}$Se$_{\rm 2}$ superconductors is a Mott insulator at ambient pressure.

We thank J. Dai, M. Fang, and T. Xiang
for useful discussions. This work was supported by
NSF Grant No. DMR-1006985 and the Robert A. Welch Foundation
Grant No. C-1411 (R.Y. and Q.S.), and U.S. DOE  at
LANL  under Contract No. DE-AC52-06NA25396 (J.-X.Z.).

\end{document}